\documentclass[a4paper,12pt]{article}

\usepackage[T2A]{fontenc}
\usepackage[cp1251]{inputenc}
\usepackage[english]{babel}
\usepackage{amssymb,amsfonts,amsmath,mathtext,cite,enumerate,float} 
\usepackage{graphicx}
\usepackage{hyperref}
\usepackage{subcaption}

\newcommand{\ptl}{\partial}

\title{Asymptotical study of two-layered discrete waveguide with a weak
coupling}
\author{A. I. Korolkov, A. V. Shanin, K. S. Kniazeva}

\begin{document}
\maketitle

\begin{abstract}
A thin two-layered waveguide is considered.
The governing equations
for this waveguide is a  matrix Klein--Gordon equation of dimension~2.
A formal solution of this system
in the form of a double integral
can be obtained by using Fourier transformation.
Then, the double integral can be reduced to a single integral with the help
of residue integration with respect to the time frequency.
However, such an integral can be difficult to estimate since it involves branching and oscillating functions. This integral is studied asymptotically.
A zone diagram technique is proposed to represent the set of possible
asymptotic formulae.
The zone diagram generalizes the
concept of far-field and near-field zones.
\end{abstract}

\section{Problem formulation}
Consider a waveguide composed of two one-dimensional layers weakly coupled with each other.
The subsystems (the layers) are described by unknown functions
$u_1(t, x)$ and $u_2(t, x)$ (here $t$ is time, and $x$ is longitudinal coordinate).
Assume that the waveguide is described by a {\em matrix Klein-Gordon equation\/}
of dimension~2
(another naming for this equation is WFEM or WaveFEM):
\begin{equation}
\label{MKGE}
\left[ \begin{pmatrix} c_1^2 \quad 0 \\ 0 \quad c_2^2 \end{pmatrix} \partial_x^2 + \begin{pmatrix} -\Omega_1^2 \qquad \mu \\ \mu \qquad -\Omega_2^2
\end{pmatrix} -
\begin{pmatrix} 1 \quad 0 \\ 0 \quad 1 \end{pmatrix} \partial_t^2 \right] \begin{pmatrix} u_1(t,x) \\ u_2(t,x) \end{pmatrix} =
\begin{pmatrix} f_1 \\ f_2 \end{pmatrix} \delta(t) \delta(x)
\end{equation}
$\begin{pmatrix} f_1 , f_2 \end{pmatrix}^T$ is a constant vector describing the excitation. The
excitation is a short pulse localized at~$x = 0$. Indeed, we are looking for a causal
solution equal to zero for~$t < 0$.
For simplicity, we take
\[
\left( \begin{array}{cc} f_1  \\ f_2 \end{array} \right) =
\left( \begin{array}{cc} 1  \\ 0 \end{array} \right).
\]

Parameter
$\mu$ is a small value describing a coupling between the subsystems.
A system with $\mu=0$ is {\em unperturbed}. Matrix coefficients
of the unperturbed system are diagonal,
and the waves in the subsystems
propagate
independently of each other.
Each subsystem is described by a scalar Klein--Gordon equation
\[
( c_j^2 \ptl_x^2 - \ptl_t^2 - \Omega_j^2) u_j (t, x) = f_j \, \delta(t) \, \delta(x).
\]
The parameters $\Omega_j$ are
cut-off frequencies of the unperturbed subsystems.

Our aim is to obtain
a formal solution of \eqref{MKGE} (see Section~\ref{FSsection})
and to get
asymptotic estimations of the wave field for different values of $t$ and~$x$.
A classical way to do this is to apply a
{\em  saddle point method\/} (see Section~\ref{SPsection}). This method allows one to describe the wave field for $x \rightarrow \infty$ and $t \rightarrow \infty$. However, for values of $x$ and $t$ that are moderately large,
one needs some more
sophisticated asymptotics. Section~\ref{ZDsection} is devoted to them.


\section{Formal solution of the matrix Klein--Gordon equation}
\label{FSsection}
\subsection{Double integral solution}

Apply Fourier transform with respect to $x$ and Laplace transform with respect to $t$ to
\eqref{MKGE}. As a result, obtain a double integral representation of the waveguide field:
\begin{equation}
\label{Double_int_repr}
\begin{pmatrix} u_1(t,x)\\u_2(t,x)\end{pmatrix} =
\int\limits_{-\infty + i\varepsilon}^{\infty + i \varepsilon}\int\limits_{-\infty}^{\infty}
\frac{{\rm A}(\omega,k)}{D(\omega,k)} e^{-i\omega t + i k x}  dk \, d\omega  ,
\end{equation}
where
\begin{equation}
\label{A_det}
{\rm A}(\omega,k)= \begin{pmatrix} -k^2 c_2^2-\Omega_2^2 + \omega^2 \\ -\mu \end{pmatrix},
\end{equation}
\begin{equation}
\label{D_det}
D(\omega,k) = (-k^2c_1^2 + \omega^2 -\Omega_1^2)(-k^2c_2^2 + \omega^2 -\Omega_2^2)-\mu^2 ,
\end{equation}
$\varepsilon$ is an arbitrary  positive real number (it provides causality of the solution).
We assume that $\epsilon$ is small.


\subsection{Series--integral solution}

Let be $x > 0$.
\eqref{Double_int_repr} can be transformed into a sum of single integrals.
For that purpose, one has to close the contour of integration with respect to $k$ in the upper half plane of $k$ (it is possible due to Jordan's lemma) and apply the residue method. As a result, obtain the following representation of the solution
represented by \eqref{Double_int_repr}:
\begin{equation}
\label{Ser_int_repr}
\begin{pmatrix} u_1(t,x)\\u_2(t,x)\end{pmatrix} = 2\pi i
\sum _{m=1}^{2}\int\limits_{-\infty + i\varepsilon}^{\infty + i \varepsilon}
\frac{{\rm A}(\omega,k_m(\omega))}{
\ptl_k D(\omega,k_m(\omega))} e^{i k_m(\omega) x - i\omega t } d\omega  ,
\end{equation}
where $\ptl_k D$ is a partial derivative of function $D$ with respect to the second argument.
Functions $k_m(\omega)$, $m = 1,2$ are two of four roots of the {\em dispersion equation}
\begin{equation}
\label{DE}
D(\omega,k)=0.
\end{equation}
solved for a fixed $\omega$.
Among four roots of this equation, we select two roots with ${\rm Im}[k_m (\omega)] > 0$
as follows.
Four roots of \eqref{DE} can be split into pairs: $\pm k_m(\omega)$, $m = 1,2$.
We denote
by $k_1(\omega)$ and $k_2(\omega)$ the solutions having positive group velocities
\[
v_{\rm gr}(\omega) \equiv \left( \partial_{\omega} k \right) ^{-1}.
\]
Solutions $k_1(\omega)$ and $k_2(\omega)$ have ${\rm Im}[k]>0$ for ${\rm Im}[\omega] = \varepsilon > 0$. This can be shown by the Taylor's series:
\begin{equation}
\label{eq0}
{\rm Im} [k(\omega' + i\varepsilon)] \approx {\rm Im}[k(\omega') + \frac{dk(\omega')}{d\omega} \cdot i\varepsilon] = \varepsilon \frac{dk(\omega')}{d\omega}, \quad \omega' = {\rm Re}[\omega].
\end{equation}


\subsection{Dispersion diagram}

A {\em dispersion diagram\/} is a graphical representation of the roots of \eqref{DE}.
We will use a real dispersion diagram, which is the set of all real points $(\omega,k)$
obeying \eqref{DE}, and a complex dispersion diagram, which is the set of complex
$(\omega,k)$ obeying \eqref{DE}.

In the real dispersion diagram,
we assign indices 1 or 2 to the branches,
depending on their asymptotics as $\omega \rightarrow \infty$.
Namely, we assume that  $k_1(\omega) \rightarrow \omega/c_1$, $k_2(\omega) \rightarrow \omega/c_2$.

\begin{figure}[!h]
\begin{center}
\includegraphics[scale=0.8]{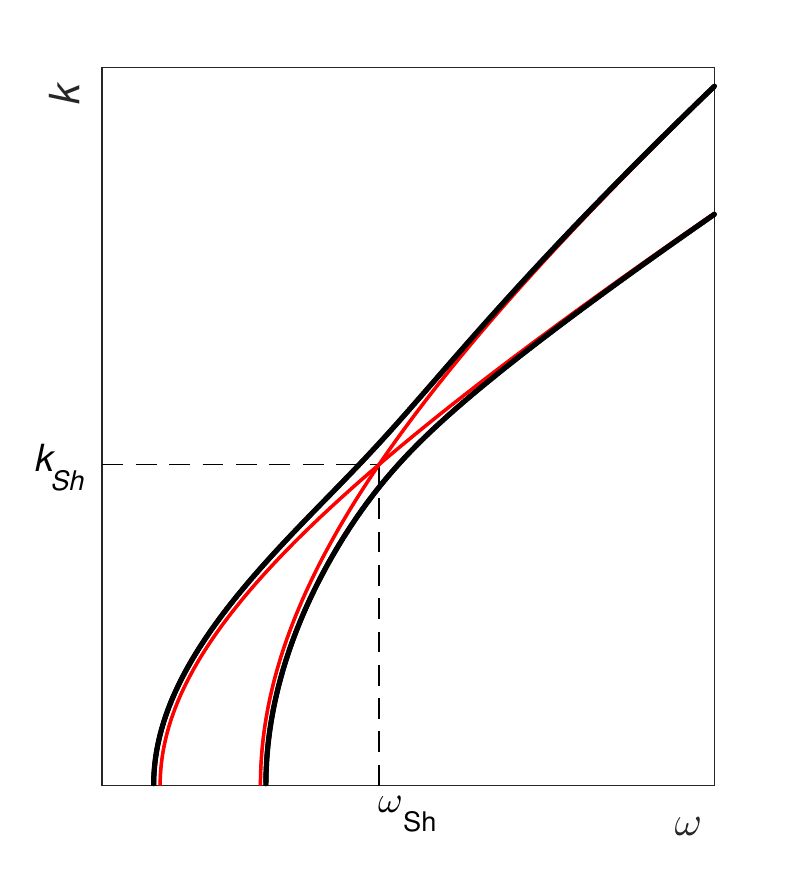}
\includegraphics[scale=0.8]{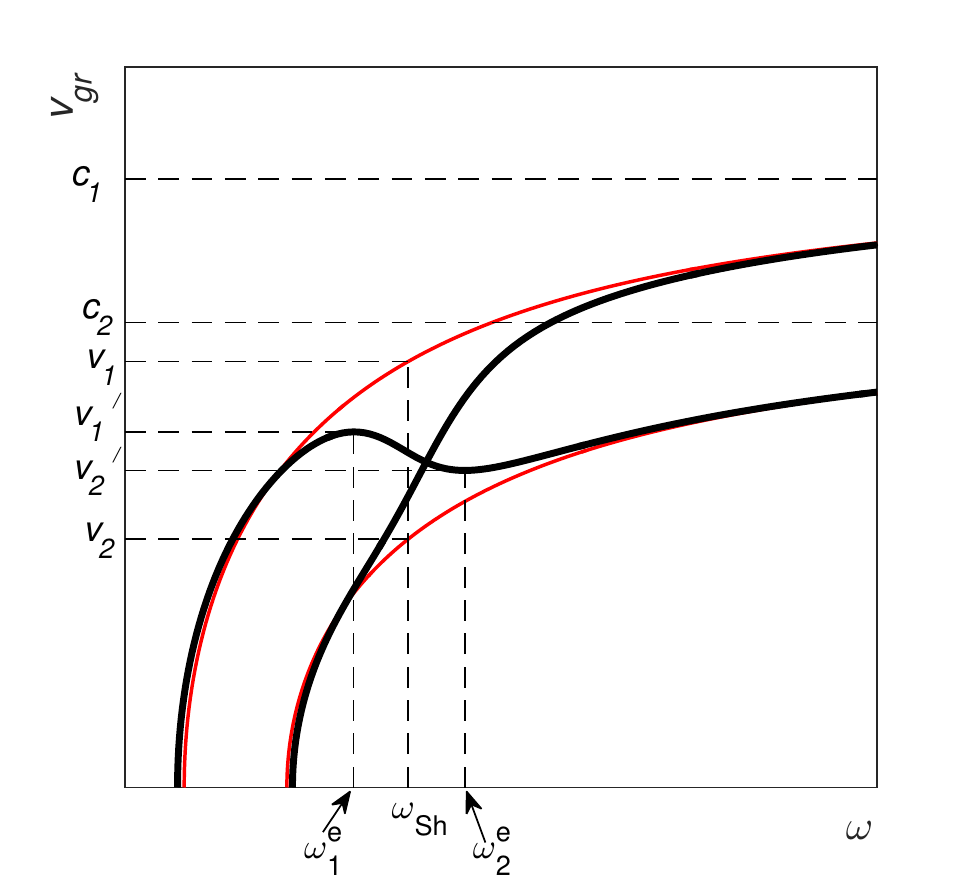}		
\end{center}
	\caption{
    Real dispersion diagram (left) and group velocities (right)
    for the system described by \eqref{MKGE}}
\label{DD}
\end{figure}

%


For a working example throughout this paper,
let us choose some values of the waveguide parameters in such a way that an {\em exchange pulse} is observed \cite{Shanin_2018}. In our case $c_1 > c_2$, $\Omega_1 <\Omega_2$. Dispersion diagram of such a system looks like in \figurename~\ref{DD}~left, black.

Dispersion diagram of the corresponding unperturbed system is
presented by
two intersecting red lines in \figurename~\ref{DD}~left. The point of intersection
$(\omega_{\rm sh},k_{\rm sh})$ is a referred to as a
Shestopalov's singular point \cite{Shestopalov1996}:
\begin{equation}
\label{Shest_Point}
\omega_{\rm sh} = \sqrt{(c_1^2 \Omega_2^2 - c_2^2 \Omega_1^2)/(c_1^2 - c_2^2)},
\qquad
k_{\rm sh} = \sqrt{(\Omega_2^2 - \Omega_1^2)/(c_1^2 - c_2^2)}.
\end{equation}
One can see that the perturbed real dispersion diagram displays an {\em avoiding crossing\/}
near the Shestopalov's point.

In \figurename~\ref{DD}~right,
the group velocities  for the system with $\mu \neq 0$ (black) and with $\mu = 0$ (red)
are displayed.
Let
$v_1'$ and $v_2'$ be a local maximum and a local minimum
of the group velocity. The corresponding frequencies are $\omega_1^e$ and $\omega_2^e$, respectively. Obviously, these points are inflection points of the real dispersion diagram.

Let $v_1$ and $v_2$ be
the values of the group velocity of the unperturbed system taken at $\omega_{\rm sh}$.
They can be easily computed:
\begin{equation}
\label{Vgr_Shest_form}
v_1 = c_1 \sqrt{\omega_{\rm sh}^2 - \Omega_{1}^2} \, / \omega_{\rm sh}, \qquad
v_2 = c_2 \sqrt{\omega_{\rm sh}^2 - \Omega_{2}^2} \, / \omega_{\rm sh}.
\end{equation}
For the systems, which we chose, $v_1 > v_2$. Obviously, $v_1 < c_1$ and $v_2 < c_2$. It can be $v_1 > c_2$ or $v_1 < c_2$ or even $v_1 = c_2$. We take $v_1 < c_2$ for definiteness.

\begin{figure}[!h]
\centering
\includegraphics[scale = 0.9]{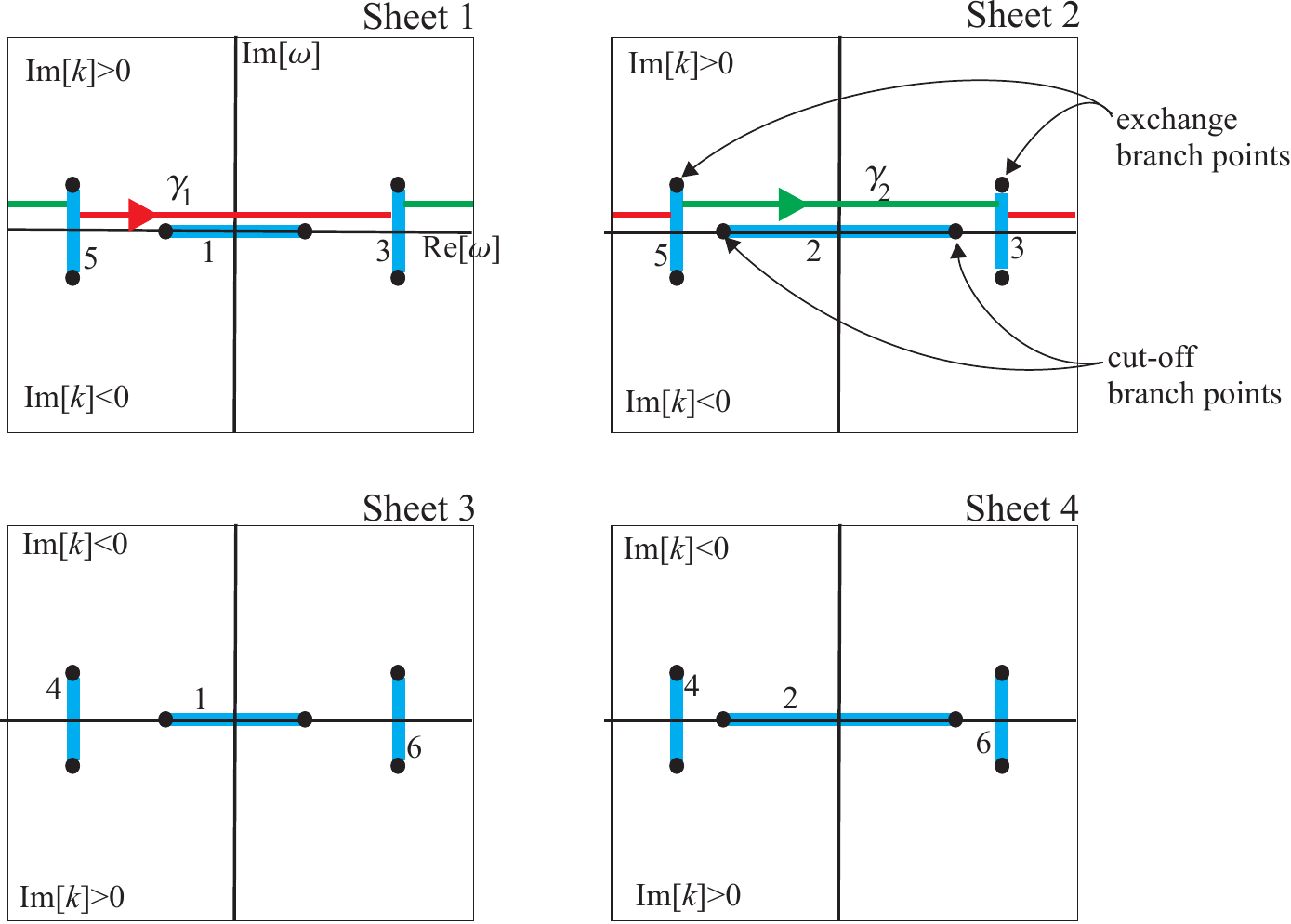}
\caption{Scheme of the Riemann surface of $k(\omega)$.
Blue lines are cuts. Equal numbers denote the sides that should be attached to each other. Integration contours $\gamma_1$ and $\gamma_2$ are the red and green lines, respectively}
\label{RS}
\end{figure}


Instead of four functions $\pm k_1(\omega)$, $\pm k_2(\omega)$, we consider a multivalued function $k(\omega)$ on the complex plane $\omega$. A scheme of Riemann surface of
the function $k(\omega)$ is shown in \figurename~\ref{RS}. This
Riemann surface has four sheets. Denote the sheet corresponding to $k_1(\omega)$ by ``Sheet~1'', the sheet for $k_2(\omega)$ by ``Sheet~2'', the sheet for $-k_1(\omega)$ by ``Sheet~3'', and the sheet for $-k_2(\omega)$ by ``Sheet~4''.

The Riemann surface has branch points of two types:
{\em cut-off branch points\/} and {\em exchange branch points}.
The cut-off branch points, are solutions of the equation
$
D(\omega,k=0) = 0.
$
These branch points
connect sheets corresponding to the solutions of the dispersion equation with opposite signs
(i.~e.\ $k_1(\omega)$ with $-k_1(\omega)$, and $k_2(\omega)$ with $-k_2(\omega)$).
The cut-off frequencies belong to the real axis of~$\omega$.

The rest of the branch points are the exchange branch points.
These points exist due to interaction between the subsystems. They connect
the sheets corresponding to different modes. In \figurename~\ref{RS} the
exchange branch points correspond to the values of $\omega$ that are not real.

One can consider \eqref{Ser_int_repr} as a contour integral of a multivalued function,
whose Riemann surface is the same of the Riemann surface of $k(\omega)$. The contour of integration is $\gamma_1 + \gamma_2$, where $\gamma_1$
and $\gamma_2$ are shown in \figurename~\ref{RS}. These contours are preimages of
the real axis of $\omega$ on Sheet~1 and Sheet~2,
slightly shifted into the upper half plane.


\section{Analysis of solutions by saddle point method (far zone)} \label{SPsection}

\subsection{Saddle point approach}

According to the Cauchy's theorem, the contour of integration
in \eqref{Ser_int_repr}
can be homotopically deformed in the area of analyticity of the integrand.
Using this fact, one can obtain some asymptotical assessments of  \eqref{Ser_int_repr}. Particularly, the saddle point method allows to estimate
the wave field in a far zone.

Introduce a {\em  formal velocity\/}  $V$ by
\[
V = x/t.
\]
Let $V$ be fixed, and let be $x \to \infty$. Indeed, this means $t \to \infty $ as well.

As it was stated above, \eqref{Ser_int_repr} can be rewritten as follows:
\begin{equation}
\label{SIR2}
u_j(t,x) = \int \limits_{\gamma_1 + \gamma_2} h_j(\omega,k(\omega)) \exp\{ ixg(\omega) \}
\, d\omega,
\quad
h_j(\omega,k) = \frac{A_j(\omega,k)}{\ptl_k D(\omega,k)},
\quad j = 1,2,
\end{equation}
where $A_j$ are components of the vector ${\rm A}$, and
\begin{equation}
\label{g_det}
g(\omega) \equiv k(\omega) - \omega / V.
\end{equation}

Since the exponential factor $\exp\{ixg\}$ contains the large parameter $x$,
while $h_j$ has not, the exponential factor mainly
determines the magnitude of the integrand. Namely,  the
integrand is {\em exponentially large\/} if ${\rm Im} [g] < 0$, and is
{\em exponentially small\/} if ${\rm Im}[g] > 0$.

Such integrals can be asymptotically estimated using the saddle point method \cite{Borovikov_1994}. Since the integration is held on the Riemann surface of a multivalued function, one has to use a multi-contour version of the saddle point method
\cite{Shanin_2018,Shanin_2017}.

According to the saddle point method,
the contour $\gamma_1 + \gamma_2$ can be deformed into a sum of several {\em saddle point contours} $\gamma_{\ast m}$:
\begin{equation}
\label{SPsum}
u_j(t,x) = \sum_m I_{j,\ast m},
\quad
I_{j,\ast m} = \int \limits_{\gamma_{\ast m}} h_j(\omega,k(\omega)) \exp\{ ixg(\omega) \}d\omega.
\end{equation}
Each saddle point contour $\gamma_{j,\ast m}$ passes through {\em a saddle point\/}
$\omega_{\ast m}$.
A saddle point $\omega_{\ast m}$ is a root of the first derivative of $g$
with an additional condition:
\begin{equation}
\label{SPdet}
\partial_{\omega} g(\omega_{\ast m}) = 0, \quad \partial_{\omega}^2 g(\omega_{\ast m}) \neq 0.
\end{equation}
After substitution of \eqref{g_det} into \eqref{SPdet} one obtains
\begin{equation}
\label{SPeq}
\frac{dk(\omega_{\ast m})}{d \omega} = V^{-1}
\quad \mbox{or} \quad
v_{\rm gr} (\omega_{\ast m}) = V.
\end{equation}

The contours $\gamma_{\ast m}$ are the contours of steepest growth of ${\rm Im}[g]$.
Consequently, the magnitude of the integrand is significant only in some vicinity of
$\omega_{\ast m}$.

Note that not all the saddle points satisfying the \eqref{SPeq} are passed by saddle point contours obtained by an appropriate deformation of $\gamma_1 + \gamma_2$.


\subsection{Wave component due to a saddle point}

Since the saddle point integration is held over some vicinity of $\omega_{\ast m}$,
$h_j(\omega,k(\omega))$ and $g(\omega)$
can be approximated by a constant $h_j(\omega_{\ast m}, k(\omega_{\ast m}))$
and by a quadratic function
\begin{equation}
\label{gTaylor}
g(\omega) \approx g(\omega_{\ast m}) + \frac{1}{2} \frac{d^2 g (\omega_{\ast m}) }{d \omega^2} (\omega - \omega_{\ast m})^2.
\end{equation}
Using \eqref{g_det} and \eqref{gTaylor},
one can obtain an approximation of the integral $I_{j,\ast m}$:
\begin{equation}
\label{SPint1}
I_{j,\ast m} = h_j(\omega_{\ast m}, k(\omega_{\ast m}))
\sqrt{\frac{2 \pi}{|\alpha| x}}
\exp\{ ik(\omega_{\ast m})x - i \omega_{\ast m} t + \mbox{sign}(\alpha) \cdot i\pi/4\}, \quad \alpha = \partial_{\omega}^2k(\omega_{\ast m}).
\end{equation}


\subsection{Merging of two saddle points near an inflection point.  Airy wave}

Consider an inflection point of the real dispersion diagram, for example
the local minimum of the group velocity (a local maximum is described in
a similar way). Such a point is shown in
\figurename~\ref{DD},~right, as the point $\omega_2^e$.
If $V$ is slightly higher than $v_2'$, there should be two real saddle points
near~$\omega_2^e$. If $V = v_2'$ these two saddle points merge with each other.
An asymptotics of wave field with $V \approx \omega_2^e$ is described by an Airy function.

Approximate function $k(\omega)$ as follows:
\begin{equation}
\label{gTaylor2}
k(\omega) \approx k(\omega^e) + \frac{\omega - \omega^e}{v'_2} - \frac{1}{3}
\alpha (\omega - \omega^e_2)^3,
\quad
\alpha = - \frac{1}{2}\partial_{\omega}^3 k(\omega_2^e),
\quad
\alpha > 0.
\end{equation}

The integral over some vicinity of $\omega^e_2$ can be approximated by an Airy function \cite{Borovikov_1994}, \cite{Brekhovskih_1980}:
\begin{equation}
\label{AiryInt1}
I_j(V, x) \approx h_j(\omega_2^e, k (\omega_2^e))\, \frac{2 \pi}{(x \alpha)^{1/3}} \,
\mbox{Ai}\left( \frac{x^{2/3}}{\alpha^{2/3}} \left(V^{-1} - (v'_2)^{-1}\right) \right) .
\end{equation}


\subsection{An overview of the saddle points  evolution}

\begin{figure}[!h]
\centering
\includegraphics[scale = 0.8]{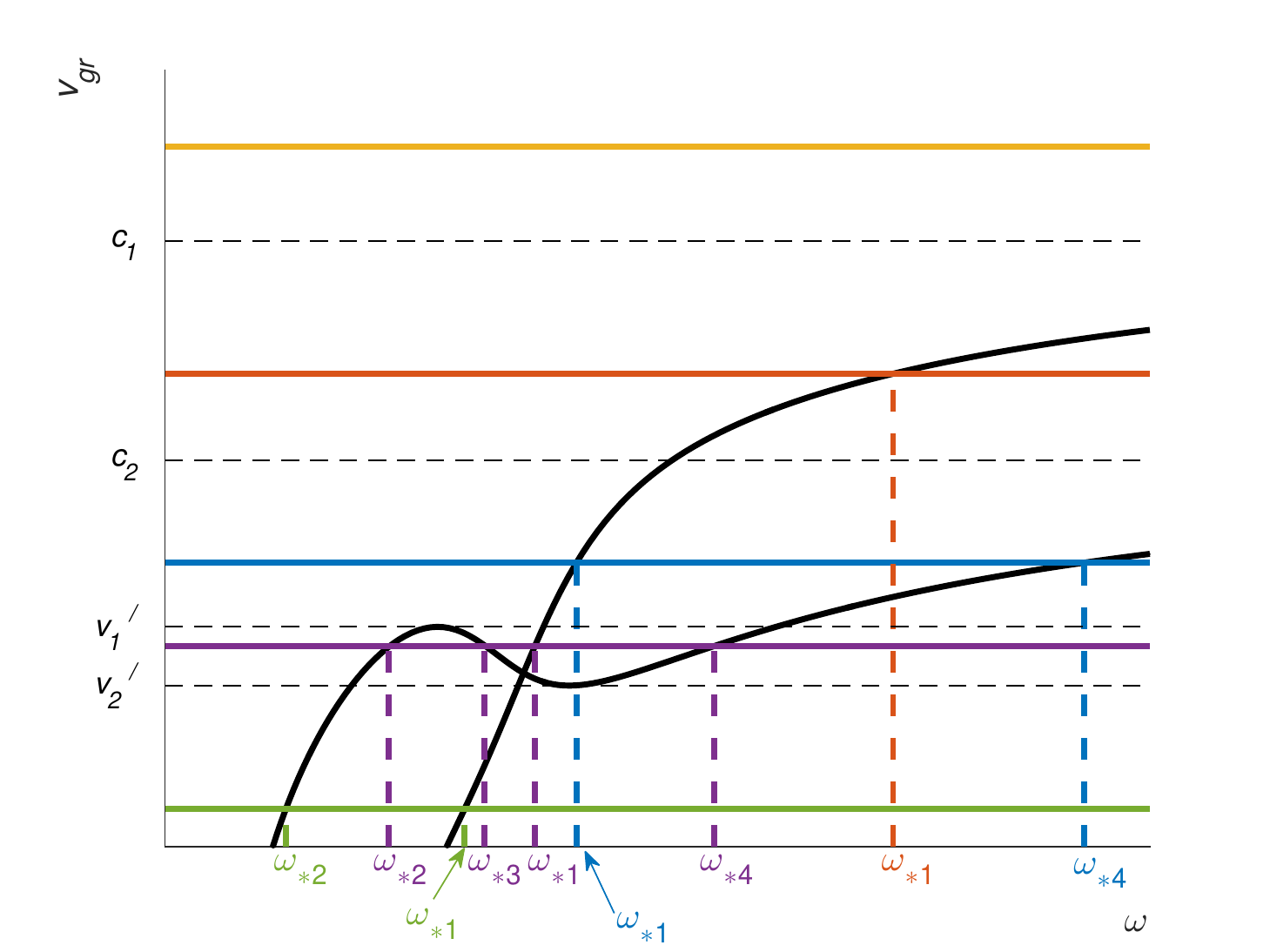}
\caption{Solving \eqref{SPeq} graphically for real $\omega$}
\label{SPnot}
\end{figure}

Let us list the {\em real\/} roots of \eqref{SPeq} for different $V$:
\begin{itemize}

\item
$V>c_1$: there are no roots. This case is shown by an orange line in \figurename~\ref{SPnot}.
Moreover, for $V>c_1 > c_2$ the wave field is equal to zero identically.
This can be shown by closing the integration contours in the upper half-plane.

\item
$c_1 > V > c_2$: there is a single real root $\omega_{\ast1}$.
The case is shown by red lines in \figurename~\ref{SPnot}.



\item
$c_2 > V > v_1'$: there are two real roots $\omega_{\ast1}$ and $\omega_{\ast4}$
shown by blue lines  in \figurename~\ref{SPnot}.

\item
$v_1' > V > v_2'$: there are four
real roots $\omega_{\ast1}$, $\omega_{\ast2}$, $\omega_{\ast3}$, $\omega_{\ast4}$.
They are shown by magenta lines in \figurename~\ref{SPnot}.

\item
$v_2' > V > 0$: there are two real roots $\omega_{\ast1}$ and $\omega_{\ast2}$. They are  shown
by  green lines in \figurename~\ref{SPnot}.
\end{itemize}

After deformation of the contour $\gamma_1 + \gamma_2$,
some
steepest descend contours pass through each of the listed saddle points.
There also exists a set of negative saddle points denoted as follows:
\[
\omega_{\ast -n}, \quad n=1,\dots,6, \quad \omega_{\ast -n} = - \omega_{\ast n}.
\]
They should be taken into account in just a similar way.

\begin{figure}[!h]
	\centering
	\begin{subfigure}{0.4\textwidth}
		\includegraphics[width=\textwidth]{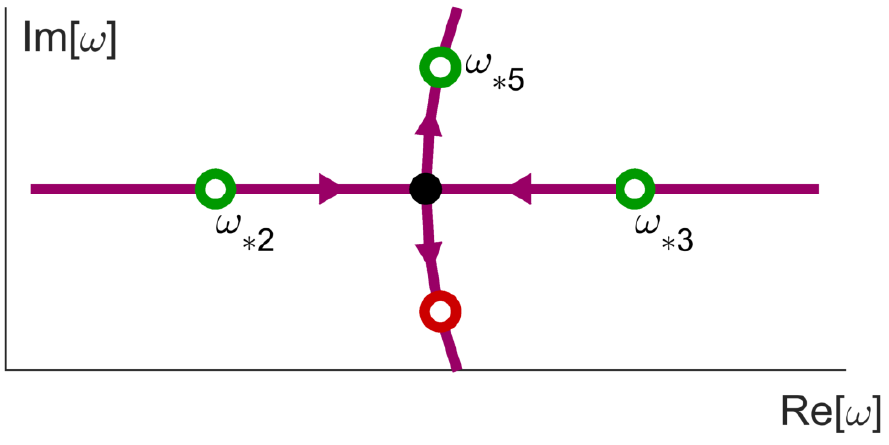}
		\caption{Vicinity of $V = v_1'$}
	\end{subfigure}
	\vspace{0.1cm}
    \hspace{8ex}
	\begin{subfigure}{0.4\textwidth}
		\includegraphics[width=\textwidth]{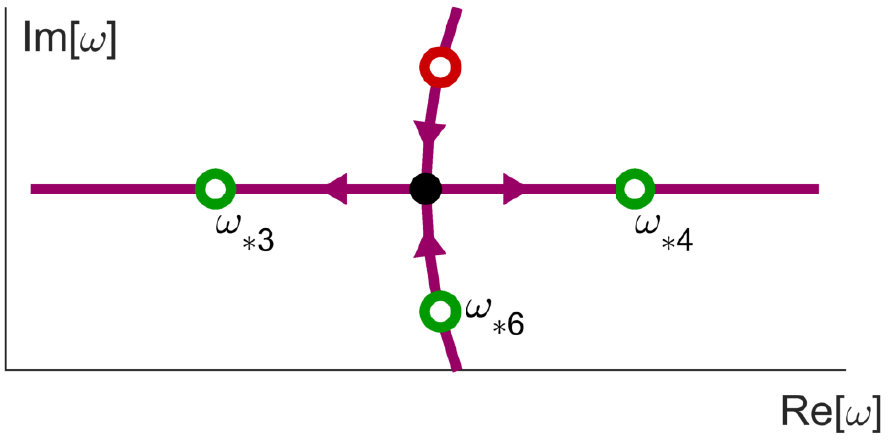}
		\caption{Vicinity of $V = v_2'$}
	\end{subfigure}
	\caption{Evolution of the saddle points in vicinities of local extrema of the group velocity. Black circles: points of an extremum, green circles: saddle points passed by steepest descend contours, red circles: saddle points not passed by steepest descend contours. The arrows
	correspond to growth of~$V$}
\label{VgrExtr}
\end{figure}

The analysis made above, together with \eqref{SPint1} and
\eqref{AiryInt1}, displays the basic structure of the field for very large~$x$.
The list contains the description of the zones, in which there exists a different number
of asymptotic terms. The formulae for the terms are given by \eqref{SPint1}.
There are also two intermediate zones ($V \approx v_1'$ and $V \approx v_2'$)
described by the Airy asymptotics of the form of \eqref{AiryInt1}.

All the wave components described above do not decay exponentially as $x$ grows.
There exist also saddle points with complex $\omega_{\ast m}$ possessing an exponential
growth. Sometimes, corresponding terms cannot be ignored (if $x$ is not large, decay is
not fast, or the decaying wave is the only wave observed for corresponding $x$ and~$t$).
Let us consider evolution of saddle points position for $V$ changing in some vicinity of $v_1'$. As $V$ increases from $v_1' - \nu$ to $v_1' + \nu$, where $\nu$ is a small real positive number, saddle points $\omega_{\ast 2}$ and $\omega_{\ast 3}$ evolve into two complex saddle points (\figurename~\ref{VgrExtr}~(a)). Among the saddle points for $V > v_1'$, only saddle point
$\omega_{\ast 5}$ is passed by deformed a contour because the initial contour of integration is shifted in the direction of growth of ${\rm Im}[g]$.

The case when $V$ grows near $v_2'$ is analogous. However, for $V < v_2'$ the deformed contour passes through the saddle point $\omega_{\ast 6}$ with negative imaginary part (\figurename~\ref{VgrExtr}~(b)).

Thus, for $v_1'<V$ and for $V<v_2'$ there may exist a complex saddle point, which gives
a decaying saddle point asymptotics.


\section{Zone diagrams} \label{ZDsection}
\subsection{The concept of zone diagram}

The saddle point method does not provide a correct asymptotics of the wave field
if $x$ is not large enough. Each saddle point has a domain of influence (DOI)
introduced in \cite{Borovikov_1994},
which is a spot on the saddle point contour on which the
integrand is not negligibly small. Typically, the size of such a spot in the
$\omega$-plane has size $\sim x^{-1/2}$. if $x$ is not very large, it may happen
that the DOI of two or more saddle points overlap. In this case the saddle point
approximation is not adequate.

An example of such a configuration is the Airy wave mentioned above, which is
a result of merging of two saddle points.
Depending on $V$ and $x$, there can be more sophisticated asymptotics.

The concept of merging of DOI of saddle points is formalized as follows.
 Let $(\omega_{\ast},k_{\ast})$ be a saddle point, and let $\gamma_{\ast}$ be the corresponding steepest descend contour. Introduce the length coordinate $l$ directed along $\gamma_{\ast}$. Let $l = 0$ correspond to the saddle point, and the coordinate direction coincide with the contour direction. The integral along $\gamma_{\ast}$ can be represented as
\begin{equation}
\label{eq1}
I_{\ast}(x) = \int \limits _{-\infty}^{\infty} h^*(l) \exp \{ ix g^*(l) \} \, dl
\end{equation}
where $h^*$ and $g^*$ are restrictions of functions
$h \, d\omega / dl$ and $g$ onto the contour.

The imaginary part of $g^*(l)$ has minimum at $l = 0$, and grows monotonically as $|l|$ grows. The DOI is a segment $[a_1,a_2]$ on the $l$-axis, such that $a_1 < 0 < a_2$, and
\begin{equation}
\label{InflDom}
x({\rm Im}[g^*(a_1)]-{\rm Im}[g^*(0)]) = x({\rm Im}[g^*(a_2)]-{\rm Im}[g^*(0)]) = {S}.
\end{equation}
Here and below,
${S}$ is some arbitrary positive moderately large real number of order 1
(a good choice is ${S} = 3$).
We find convenient to write $a > S$ instead of $a \gg 1$.

The segment $[a_1,a_2]$ is, indeed, just an estimation of the DOI.
One can see that
the size of the DOI depends on $x$, and the domain becomes smaller as $x$ grows.

One needs an Airy representation (\eqref{AiryInt1})
when the DOIs of the saddle points $\omega_{\ast 2}$ and
$\omega_{\ast 3}$  (or $\omega_{\ast 3}$ and $\omega_{\ast 4}$) overlap. Note that the Airy function ${\rm Ai}(z)$ has two asymptotics: for $z \gg 1$ and for $-z \gg 1$. One of these asymptotics correspond to two real saddle points, while another corresponds to a single
complex saddle point.

Evaluate the phase difference for any pair of
{\em neighbouring} saddle points $\omega_{\ast m}$ and $\omega_{\ast n}$:
\begin{equation}
\label{Phase_deff}
\Delta_{m,n} \equiv |(k(\omega_{\ast m})x - \omega_{\ast m} t) -
(k(\omega_{\ast n})x - \omega_{\ast n} t) |
=
x\, |\, g^*(\omega_{\ast m}) -  g^*(\omega_{\ast n}) \,|.
\end{equation}
If $\Delta_{m,n}$ is bigger than ${S}$ then
$\omega_{\ast m}$ and $\omega_{\ast n}$ can be considered as separate saddle points.
If  $\Delta_{m,n} < {S}$
then one should introduce some more sophisticated asymptotics.

Let us say when we consider saddle points $\omega_{\ast m}$ and $\omega_{\ast n}$ being neighboring. This is when there exists a path between $\omega_{\ast m}$ and $\omega_{\ast n}$ along the Riemann surface of $k(\omega)$, on which the phase $xg(\omega)$
does not change strongly (for more than ${S}$).


The asymptotics of a waveguide can be organized in {\em a zone diagram}.
We plot the zone diagram in the coordinates $(t,V) = (t , x/t)$,
The diagram is branched, i.~e.\ over each point $(t, V)$ there may exist several sheets
of the zone diagram. Each sheet corresponds to a separate asymptotic term.

There are parent / child relations between the asymptotics on the diagram. The
``children'' are obtained by taking some asymptotics of the ``parents''.


\subsection{A simple example: a zone diagram for a scalar Klein--Gordon equation}

Consider a scalar inhomogeneous Klein--Gordon equation
\begin{equation}
(c^2 \ptl_x^2 - \ptl_t^2 - \Omega^2) u(t, x) = \delta(t) \, \delta(x).
\label{SKGE}
\end{equation}
A causal solution of this equation can be easily obtained:
\begin{equation}
u(t, x) = \left\{ \begin{array}{ll}
-(2c)^{-1} J_0 (\Omega \sqrt{t^2 - x^2 / c^2}) & t > x /c \\
0                                              & 0 < t < x /c
\end{array} \right.
\label{SKGEsol}
\end{equation}
where $J_0$ is the Bessel function.

One can see that there are two asymptotics of the solution:
if the argument of the Bessel function is much greater than~1 (we prefer to say that it is greater than
${S}$), then one can represent the Bessel function in terms of two exponential functions:
\begin{equation}
u(t, x) \approx -\frac{1}{2c \sqrt{2 \pi z}}
(\exp \{i z - i \pi / 4 \} + \exp \{-i z + i \pi / 4 \}) ,
\quad
z = \Omega \sqrt{t^2 - x^2 / c^2},
\quad
z > {S}.
\label{SKGEas1}
\end{equation}
This is the ``far field'' for the waveguide.

If the argument is small, there is another asymptotics:
\begin{equation}
u(t, x) \approx - (2c)^{-1},
\quad
z < {S}^{-1}.
\label{SKGEas2}
\end{equation}
This is the ``near field''.

\begin{figure}[!h]
	\centering
	\begin{subfigure}{0.4\textwidth}
		\includegraphics[width=0.8 \textwidth]{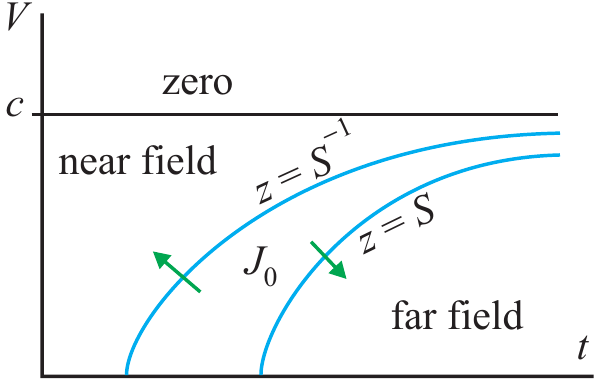}
		\caption{Zone diagram in the $(t, V)$-plane}
	\end{subfigure}
	\vspace{0.1cm}
    \hspace{8ex}
	\begin{subfigure}{0.4\textwidth}
		\includegraphics[width=0.8 \textwidth]{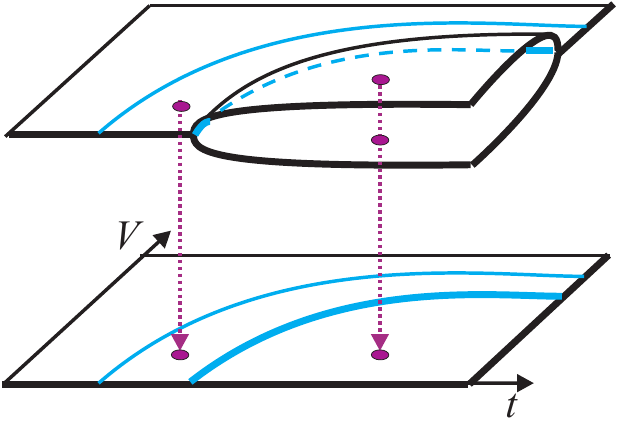}
		\caption{Branching of the diagram}
	\end{subfigure}
	\caption{Zone diagram for the scalar Klein--Gordon equation}
\label{SKGEdiagram}
\end{figure}

A zone diagram for the scalar Klein--Gordon equation is plot in
\figurename~\ref{SKGEdiagram}. The left part of the figure is the diagram projected onto
the $(t, V)$-plane, and the left part shows the branching of the diagram.
One can see that the zone of the diagram labeled as ``far field'' is represented by two sheets.
These sheets denote two terms of \eqref{SKGEas1}. Indeed, these terms can be treated as
the saddle point contributions of a well-known integral representation
for the Bessel function.

The central part of the diagram
denoted by $J_0$ is the domain where one cannot use any asymptotics, and should use the
initial expression given by \eqref{SKGEsol}.
In the ``near field'' zone the field is the constant given by \eqref{SKGEas2}.

The green arrows show the parent / child relations between the zones.


\subsection{Zone diagram for 2D matrix Klein--Gordon equation}

Building of a zone diagram for the matrix Klein--Gordon equation (\eqref{MKGE}) is a  sophisticated task. The reason is that there are four saddle points
($\omega_{\ast 1} , \dots , \omega_{\ast 4}$) that can merge.
As a result, the zone diagram becomes complicated. Here we do not build the
complete zone diagram, namely, we don't show the near field zones.

\begin{figure}[!h]
\centering
\includegraphics[scale = 1.1]{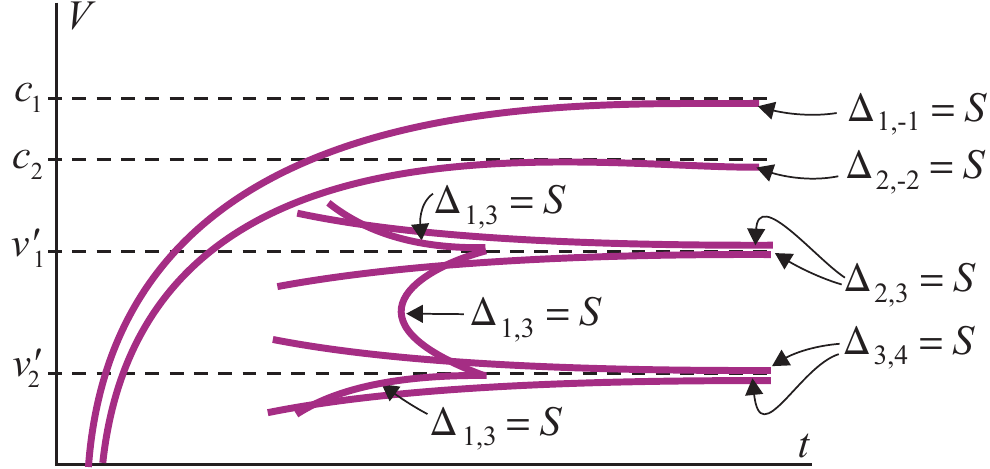}
\caption{Boundaries between zones for \eqref{MKGE}}
\label{ZD1}
\end{figure}

A sketch of the boundaries between the zones is shown in \figurename~\ref{ZD1}.
Corresponding DOIs overlap if the point $(t , V)$ is located to the left of a boundary.

\begin{figure}[!h]
\centering
\includegraphics[scale = 1]{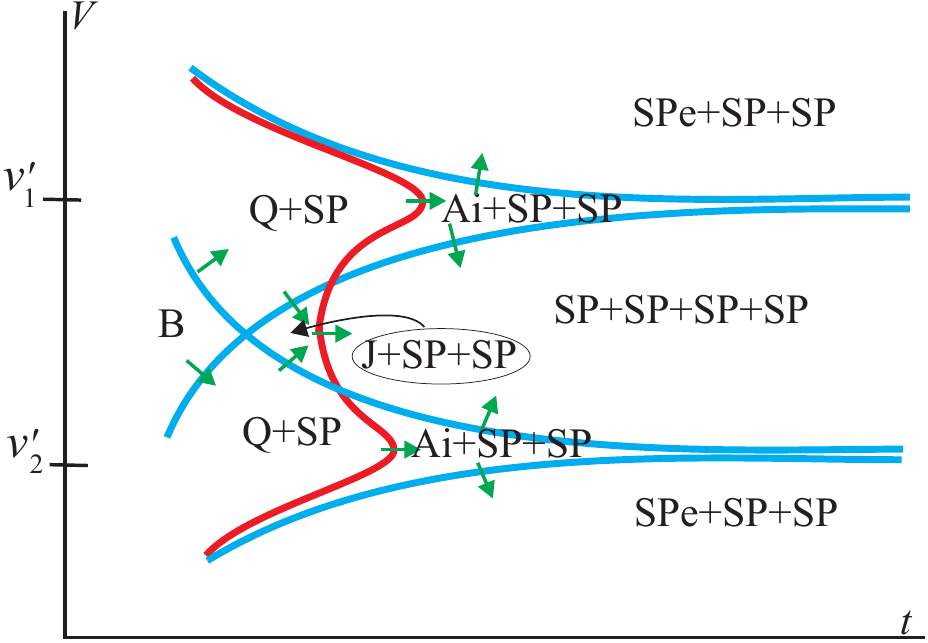}
\caption{Central part of the zone diagram}
\label{ZD2}
\end{figure}

The central part of the zone diagram is shown in more details
in \figurename~\ref{ZD2}.
The letter notations describe the asymptotic terms that exist
in corresponding zones.

Note that for the sake of compact notations, only the terms produced by
saddle points located in the right half-plane of $\omega$ are indicated.
In a complete description of terms, our notations should be doubled.

Let us list the types of the asymptotics.

{\bf SP}: A term due to a real saddle point (see \eqref{SPint1}).

{\bf SPe}: A term due to a saddle point with complex $\omega_{\ast m}$. Such a term has
a structure close to  \eqref{SPint1}, but the exponential factors decays as $x$ grows.

{\bf Ai}: A term having Airy-type asymptotics of the form (\eqref{AiryInt1}).
The diagram indicates that the ``Ai'' term degenerates into
two ``SP'' terms or into a single ``SPe'' term.

\begin{figure}[!h]
\centering
\includegraphics[scale = 1]{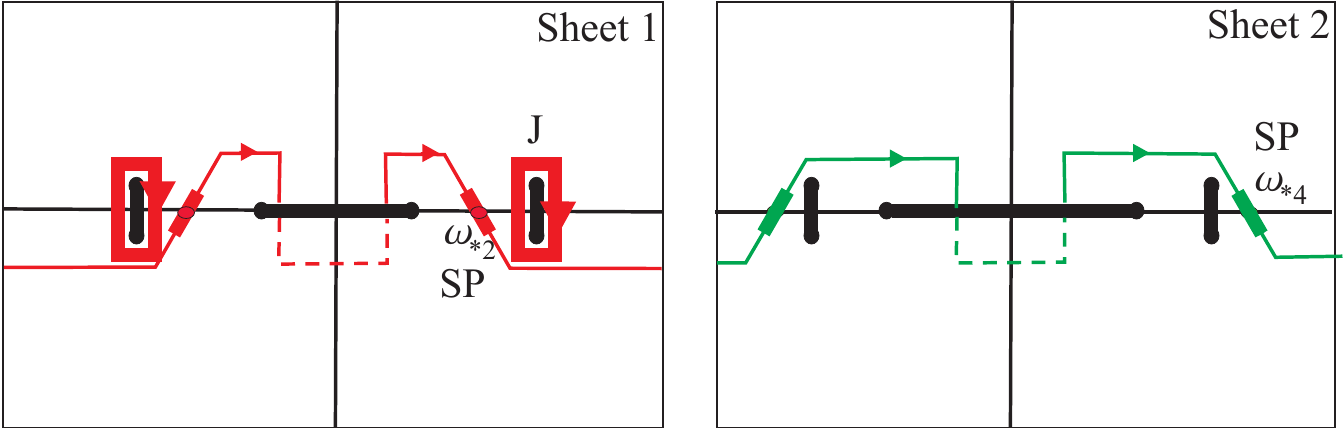}
\caption{The deformed contour of integration for the points
$(t, V)$, for which the asymptotics J exists}
\label{cont_J}
\end{figure}

{\bf J}: Such a term can be observed in the zone where the DOIs of the saddle points
$\omega_{\ast 1}$ and $\omega_{\ast 3}$ merge, while the saddle points
$\omega_{\ast 2}$ and $\omega_{\ast 4}$ are isolated.
\figurename~\ref{cont_J}
shows the integration contour $\gamma_1 + \gamma_2$
after an appropriate deformation
corresponding to this point. Bold parts of the contours show the points of the complex plane
$\omega$, where the magnitude of the integrand is not exponentially small. The points
$\omega_{\ast 2}$ and $\omega_{\ast 4}$ produce standard saddle point terms.
Another significant part of the integral is computed  along the contours encircling the cuts. This term is the J asymptotics.
We should note that this term
describes the exchange pulse \cite{Shanin_2018}.

Let us compute the J-type asymptotics of the field.
Introduce new coordinates
\begin{equation}
\label{new_var}
\omega' = \omega - \omega_{\rm sh}, \quad k' = k - k_{\rm sh}.
\end{equation}
We assume that
\[
\frac{k'}{k_{\rm sh}} \sim \frac{\omega'}{\omega_{\rm sh}} << 1.
\]
Applying the perturbation method one can obtain the following representation of the dispersion function $D(\omega,k)$ (\eqref{D_det}):
\begin{equation}
\label{D_as}
D \approx 4c_1^2 c_2^2 k_{\rm sh}^2 \left( k' - \frac{\omega'}{v_1} \right)\left( k' - \frac{\omega'}{v_2} \right) - \mu^2.
\end{equation}
Substituting \eqref{D_as} into \eqref{DE} one can obtain the following representation of the function $k(\omega)$:
\begin{equation}
\label{k_asympt}
k(\omega) \approx k_{\rm sh} + \frac{v_1^{-1} + v_2^{-1}}{2}\omega' \pm \sqrt{\frac{(v_2^{-1} - v_1^{-1})^2 (\omega')^2}{4} + \frac{\mu^2}{4c_1^2 c_2^2 k_{\rm sh}^2}}.
\end{equation}
Introduce a stretched variable
\begin{equation}
\label{coord}
\tau = \xi \omega' ,
\end{equation}
where
\[
\xi = \frac{c_1 c_2 k_{\rm sh} (v_2^{-1} - v_1^{-1})}{\mu}.
\]
The
J-asymptotics can be expressed as an integral
\begin{equation}
\label{J_int}
I_j \approx -\frac{i \pi}{2 c_1^2 c_2^2 k_{\rm sh}^2 (v_2^{-1} - v_1^{-1})}
A_j(\omega_{\rm sh},k_{\rm sh})
\,
\exp\{ ik_{\rm sh}x -i \omega_{\rm sh} t  \}
\,\times
\qquad \qquad \qquad \qquad \qquad \qquad \qquad \qquad
\end{equation}
\[
\qquad \qquad \qquad \qquad \qquad
\int \limits_{\sigma}
\exp\left\{
-i\frac{\tau}{\xi} \left( t - x\frac{ (v_1^{-1} + v_2^{-1})}{2} \right) +
i  \frac{x\, \mu \sqrt{1+\tau^2}}{2 c_1 c_2 k_{\rm sh}}
\right\}
\frac{d\tau}{\sqrt{1 + \tau^2}}
,
\]
where $\sigma$ is the contour encircling the cut $[-i,i]$ in the positive direction.
This integral can be expressed in terms of Bessel function $J_0$:
\begin{equation}
\label{Jasympt}
I_j = -\frac{\pi^2}{c_1^2c_2^2 k_{\rm sh}^2
(v_2^{-1} - v_1^{-1}) }
A_j(\omega_{\rm sh},k_{\rm sh}) \exp\{ik_{\rm sh}x -i \omega_{\rm sh} t \}
J_0(b),
\end{equation}
\[
b = \frac{ \mu \sqrt{(t - x / v_1)(x / v_2 - t)}}{c_1 c_2 k_{\rm sh} (v_2^{-1} - v_1^{-1})}
\]

{\bf Q}: This asymptotics is valid when the DOIs of the saddle points
$\omega_{\ast 1}$, $\omega_{\ast 2}$, and $\omega_{\ast 3}$
overlap (or when the DOIs
of $\omega_{\ast 1}$, $\omega_{\ast 3}$, and $\omega_{\ast 4}$ overlap). Although we don't present an explicit expression for the Q-asymptotics here, we describe a way to obtain it.

\begin{figure}[!h]
\centering
\includegraphics[scale = 1]{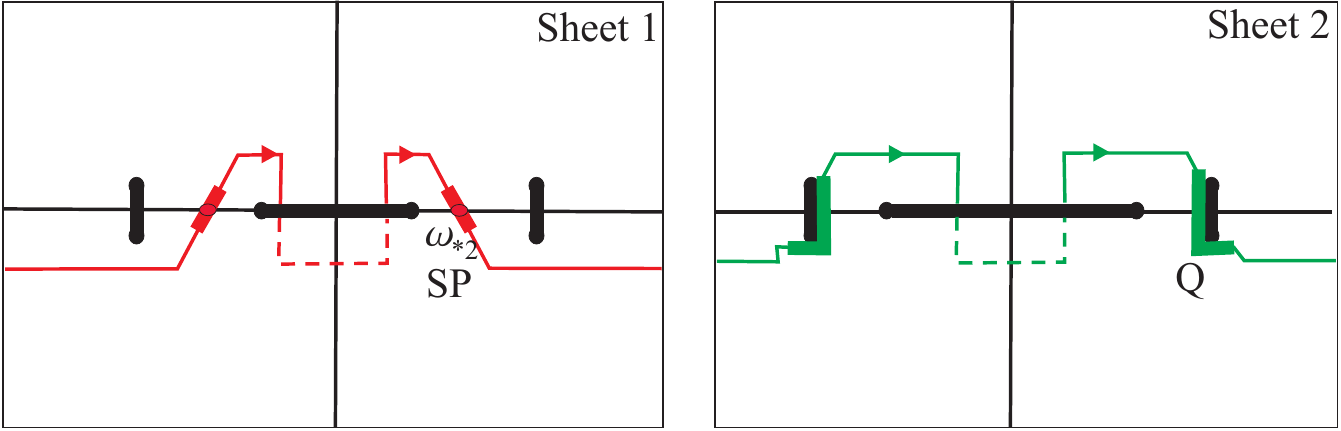}
\caption{The deformed contour of integration for the points
$(t, V)$ with the asymptotics Q}
\label{cont_Q}
\end{figure}

In \figurename~\ref{cont_Q} we plot deformed contours of integration for
the lower domain with the Q-asymptotics
when the DOIs of $\omega_{\ast 1}$, $\omega_{\ast 3}$, and $\omega_{\ast 4}$ overlap,
and $\omega_{\rm ast 2}$ is isolated.
The bold green contour on Sheet~2 going along the cut corresponds to the Q-asymptotics.

%

\begin{figure}[!h]
\centering
\includegraphics[scale = 0.9]{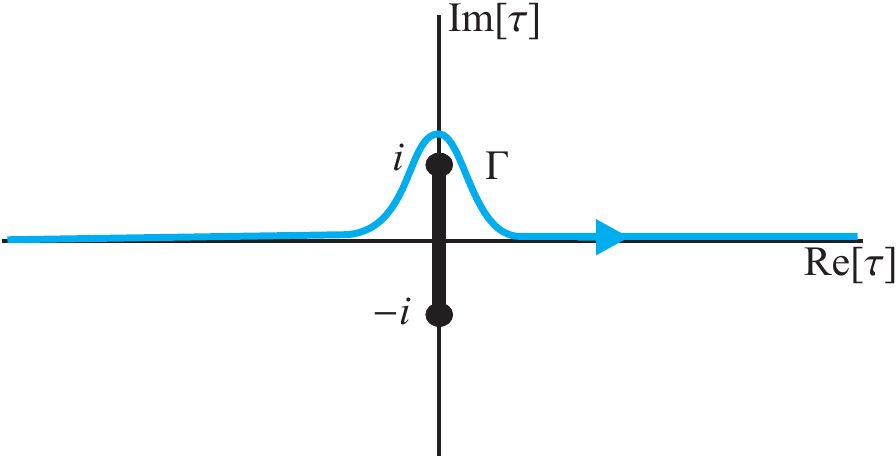}
\caption{Contour $\Gamma$ for defenition of $Q(\beta,z)$}
\label{cont_gamma_Q}
\end{figure}

Some detailed consideration shows that
an
 approximation of the field corresponding to the Q-asymptotics
  can be expressed through a new special function
\begin{equation}
\label{asymptQ}
Q(\beta,z) =  \int \limits_{\Gamma} \frac{1}{\sqrt{1 + \tau^2}}
\exp\left\{i (\sqrt{1 + \tau^2} + z\tau + \beta \tau^2)\right\} d \tau,
\end{equation}
where $\beta$, and $z$ are some constants, which depend on the waveguide parameters,
contour $\Gamma$ is shown in \figurename~\ref{cont_gamma_Q}.

{\bf B}: In this case the DOIs of all saddle points overlap. This zone is the parent for all, and
no simplification of the integral
\eqref{Ser_int_repr} can be made.


\section{Conclusion}

The saddle point method allows one to obtain asymptotic
estimations of the field in the far zone.
This zone is the set of points $(t, x)$,
such  that the
domains of influence of different saddle points do not overlap.

If the domains of influences of different saddle points overlap, one has to build new asymptotics. A parent / child relation between the new and old asymptotics
can be established.

To visualize all the obtained asymptotics, we propose a zone diagram.
This is a branched diagram in the $(t, V)$-plane.

In this work we built such set of asymptotics for a system described by \eqref{MKGE}.
A formula for the J-asymptotics is derived.
A sketch of obtaining a Q-asymptotics is given.

\section{Acknowledgements}
The work is supported by the RFBR grant 19-29-06048.

\bibliography{biblio}
\bibliographystyle{ieeetr}

\end{document}